%% file: main.tex
\documentclass[11pt]{article}

\usepackage[T1]{fontenc}
\usepackage[utf8]{inputenc}
\usepackage[english]{babel}
\usepackage{lmodern}
\usepackage{microtype}
\usepackage[margin=1in]{geometry}
\usepackage{amsmath,amssymb,amsthm,bm}
\usepackage{mathtools}
\usepackage{graphicx}
\usepackage{booktabs}
\usepackage{tabularx}
\usepackage{array}
\usepackage{caption}
\usepackage[section]{placeins}
\usepackage{subcaption}
\usepackage{enumitem}
\usepackage{xcolor}
\usepackage{listings}
\usepackage{tikz}
\usetikzlibrary{arrows.meta,positioning,fit}
\usepackage[numbers,sort&compress]{natbib}
\usepackage[colorlinks=true,citecolor=blue!60!black,linkcolor=blue!60!black,
            urlcolor=blue!60!black]{hyperref}

\setlist{itemsep=0.25ex,topsep=0.5ex}
\newcommand{\pkg}[1]{\texttt{#1}}
\newcommand{\code}[1]{\texttt{#1}}
\newcommand{\T}{^{\top}}
\newcommand{\yl}{\bm{y}_l}
\newcommand{\y}{\bm{y}}
\newcommand{\yhat}{\hat{\bm{y}}}
\newcommand{\X}{\bm{X}}
\newcommand{\Xl}{\bm{X}_l}
\newcommand{\C}{\bm{C}}
\newcommand{\Sig}{\bm{\Sigma}}
\newcommand{\V}{\bm{V}}
\newcommand{\bhat}{\hat{\bm{\beta}}}
\newcommand{\bb}{\bm{\beta}}
\newcommand{\uhat}{\hat{\bm{u}}_l}
\newcommand{\I}{\bm{I}}
\newcommand{\Lag}{\bm{L}}
\DeclareMathOperator{\diag}{diag}
\DeclareMathOperator{\Var}{Var}
\DeclareMathOperator{\rss}{rss}

\definecolor{codebg}{HTML}{F6F8FA}
\definecolor{codekw}{HTML}{0B4F8A}
\definecolor{codestr}{HTML}{0A6E55}
\definecolor{codecom}{HTML}{6A737D}
\lstdefinestyle{python}{
  language=Python,
  backgroundcolor=\color{codebg},
  basicstyle=\ttfamily\footnotesize,
  keywordstyle=\color{codekw}\bfseries,
  stringstyle=\color{codestr},
  commentstyle=\color{codecom}\itshape,
  showstringspaces=false,
  columns=fullflexible,
  keepspaces=true,
  breaklines=true,
  frame=single,
  rulecolor=\color{black!15},
  framesep=4pt,
  xleftmargin=4pt,
  xrightmargin=4pt,
  captionpos=b,
  upquote=true,
  literate={≈}{{$\approx$}}1,
}
\lstdefinestyle{output}{
  basicstyle=\ttfamily\scriptsize,
  backgroundcolor=\color{white},
  frame=single,
  rulecolor=\color{black!15},
  columns=fixed,
  keepspaces=true,
  showstringspaces=false,
  captionpos=b,
  upquote=true,
}
\title{\pkg{tempdisagg}: A Python Framework for Temporal Disaggregation\\
       of Time Series Data}
\author{Jaime Vera-Jaramillo\thanks{Email:
        \href{mailto:jaimevera1107@gmail.com}{jaimevera1107@gmail.com},
        \href{mailto:ja.veraj@uniandes.edu.co}{ja.veraj@uniandes.edu.co};
        ORCID: \href{https://orcid.org/0000-0001-7611-6251}{0000-0001-7611-6251}}}
\date{}

\begin{document}
\maketitle

\begin{abstract}
\noindent
\pkg{tempdisagg} is an open-source Python package that estimates a
high-frequency time series from low-frequency benchmarks and, optionally,
high-frequency indicators, such that the estimates aggregate exactly to the
benchmarks. It implements the regression-based estimators of Chow--Lin,
Fernández and Litterman, the dynamic Chow--Lin model, the Denton and
Denton--Cholette benchmarking procedures and uniform distribution, with the
formulas and defaults of the R package \pkg{tempdisagg}, together with the
autoregressive distributed-lag model of Proietti as an experimental extension.
All generalised least squares estimators share one engine in which the
residual covariance is applied through its triangular factor, so that one
likelihood evaluation costs $O(n\,n_l^2)$ instead of $O(n^2 n_l)$. The package
reports complete inference (coefficients, standard errors, likelihood-based
criteria, prediction variances and a full profile of the autoregressive
parameter with identification diagnostics), accepts additional exact linear
constraints, multiple indicators and aggregation weights, and provides exact
non-negative and cross-sectional reconciliation, rolling-origin backtesting,
revision analysis, conformity diagnostics and a common-factor test. Agreement
with R is enforced by an automated test suite over 29 reference cases:
with a fixed autoregressive parameter predictions and statistics agree to
relative differences between $10^{-15}$ and $10^{-7}$, and with an estimated
parameter they agree to about $10^{-8}$ once the difference in optimiser tolerance
is removed. Experiments with series whose high-frequency values are known
illustrate how the reported tools support the choice of method.
\end{abstract}


\input{sections/introduction}
\input{sections/framework}
\input{sections/preprocessing}
\input{sections/assessment}
\input{sections/software}
\input{sections/validation}
\input{sections/discussion}

\section*{Code and data availability}
\begin{itemize}
  \item Source code: \url{https://github.com/jaimevera1107/tempdisagg} (MIT license).
  \item Distribution: \url{https://pypi.org/project/tempdisagg/}; this paper
        describes version 0.6.6.
  \item Requirements: Python $\geq 3.10$, NumPy, pandas and SciPy; matplotlib
        is optional.
  \item Data: the Swiss pharmaceutical data are distributed with the package
        (\code{load\_swisspharma} in \code{tempdisagg.datasets}); the
        U.S.\ macroeconomic data are distributed with \pkg{statsmodels}
        (data set \code{macrodata}).
  \item Reproducibility: the script \code{reproduce\_paper.py}, distributed
        as an ancillary file of this paper and in the \code{examples} folder of
        the repository, regenerates every table, figure and number of
        Section~\ref{sec:validation}.
\end{itemize}

\section*{Acknowledgments}
The statistical design follows the R package \pkg{tempdisagg} by Christoph Sax
and Peter Steiner, which also serves as the numerical reference of this work.
The package builds on NumPy~\citep{harris2020numpy}, SciPy~\citep{virtanen2020scipy}
and pandas~\citep{mckinney2010pandas}.

\section*{Funding and competing interests}
This work received no external funding. The author declares no competing
interests.

\bibliographystyle{plainnat}
\bibliography{references}

\end{document}

%% file: sections/introduction.tex
\section{Introduction}\label{sec:introduction}

Many economic and social statistics are compiled at a lower frequency than the
one at which they are needed. National accounts are benchmarked to annual
surveys, labour-market and population figures are published annually or
quarterly, and administrative registers are often consolidated once a year.
Temporal disaggregation methods estimate the unobserved high-frequency path of
such a series---quarterly from annual, monthly from quarterly, daily from
monthly---using the low-frequency values as exact constraints and, when
available, related high-frequency indicators as information about the
short-term movement \citep{chow1971best,denton1971adjustment,dagum2006benchmarking}.
The constraint distinguishes the problem from interpolation or forecasting:
the estimate must reproduce the published totals, averages, or first or last
values exactly, so that the high- and low-frequency figures are mutually
consistent.

Two families of methods dominate practice and the guidelines of statistical
offices \citep{bloem2001qna,eurostat2018guidelines}. Regression-based methods
\citep{chow1971best,fernandez1981methodological,litterman1983random,santossilva2001chowlin}
relate the target to the indicators through a linear model estimated by
generalised least squares (GLS) on the aggregated data, and distribute the
low-frequency residuals according to an assumed high-frequency residual
process. Benchmarking methods \citep{denton1971adjustment,dagum2006benchmarking}
take one indicator as a preliminary series and adjust it as little as possible,
in a quadratic sense, to satisfy the constraints. The R package
\pkg{tempdisagg} \citep{sax2013temporal} implements both families with a common
interface and has become a reference implementation. Practical experience with
these methods is summarised by \citet{quilis2018temporal}, who emphasises that
the estimates reproduce the benchmarks by construction, so that in-sample fit
does not rank them, and that the choice of method should rest on
diagnostics such as revisions, the behaviour of the autoregressive parameter
and conformity with the indicators.

This paper describes \pkg{tempdisagg}, a Python package for temporal
disaggregation. Its design is guided by three requirements. First,
\emph{equivalence}: for every method shared with the R package, estimates and
reported statistics must coincide with those of R up to numerical error, and
this is verified automatically. Second, \emph{transparency}: the package does
not impute inputs, fall back silently to another method or use pseudo-inverses;
artificial values introduced by preprocessing are recorded, and statistical
problems such as a flat likelihood in the autoregressive parameter, an optimum
on the boundary of the search interval or competing local optima are reported
as typed warnings and stored with the result. Third, \emph{assessment}: since
exact aggregation makes in-sample comparisons uninformative, the package
provides out-of-sample and real-time tools to compare candidate models.

The contributions of the package are:
\begin{itemize}
  \item A single GLS engine for all regression-based methods, in which each
    method is a residual structure (Section~\ref{sec:framework}). The residual
    covariance matrix is never formed: products with its triangular factor are
    computed by recursive filters, which reduces the cost of a likelihood
    evaluation from $O(n^2 n_l)$ to $O(n\,n_l^2)$ for $n$ high-frequency and
    $n_l$ low-frequency observations.
  \item Estimation of the autoregressive parameter by maximum likelihood or
    minimum weighted residual sum of squares over the whole admissible
    interval, with the complete profile, truncation information and
    identification diagnostics returned to the user.
  \item Complete inference: coefficient tests, likelihood-based information
    criteria, and the variance of the prediction error of the best linear
    unbiased predictor \citep{goldberger1962best}, from which pointwise
    prediction intervals are computed.
  \item Exact sparse solutions of the Denton and Denton--Cholette problems with
    an explicit check of the identification condition.
  \item Additional information as exact linear constraints (known
    high-frequency values, benchmarks at intermediate frequencies), aggregation
    weights for rates, exact non-negative projection and two-step
    cross-sectional reconciliation \citep{difonzo2011simultaneous}.
  \item Assessment tools: rolling-origin backtesting, real-time revision
    analysis and conformity diagnostics in the sense of
    \citet{quilis2018temporal}, and a test of the common-factor restriction that
    nests Chow--Lin in the autoregressive distributed-lag model of
    \citet{proietti2006temporal}.
  \item A test suite that compares 29 configurations with outputs generated by
    the R package, checks every optimised computation against a dense textbook
    implementation, and verifies statistical properties by simulation.
\end{itemize}

The remainder of the paper is organised as follows.
Section~\ref{sec:framework} presents the statistical framework and the
estimators. Section~\ref{sec:preprocessing} describes the treatment of
irregular inputs, additional constraints and reconciliation.
Section~\ref{sec:assessment} presents the assessment tools.
Section~\ref{sec:software} describes the software design and its testing.
Section~\ref{sec:validation} reports the validation study, and
Section~\ref{sec:discussion} discusses limitations and concludes.

%% file: sections/framework.tex
\section{Statistical framework and estimators}\label{sec:framework}

\subsection{Aggregation constraint}

Let $\y = (y_1,\dots,y_n)\T$ denote the unknown high-frequency series and
$\yl$ the $n_l$ observed low-frequency values. Each low-frequency period $g$
contains a set of consecutive high-frequency periods, and the relation between
them is linear,
\begin{equation}\label{eq:constraint}
  \C\y = \yl ,
\end{equation}
where the aggregation matrix $\C \in \mathbb{R}^{n_l\times n}$ has in row $g$
the weights of the sub-periods of $g$ and zeros elsewhere. The package supports
four conversion rules (Table~\ref{tab:conversion}). The number of sub-periods
$m_g$ may vary across periods, which matters for calendar frequencies such as
days within months. High-frequency periods that belong to no observed
low-frequency period have a zero column in $\C$; they are estimated by the
same model and reported as extrapolated or retropolated values
\citep{wei2006time}. A low-frequency period whose value is missing has no row
in $\C$: its sub-periods are estimated without a constraint, and no value is
imputed for it.

\begin{table}[!htbp]
  \centering\small
  \caption{Conversion rules: weights of row $g$ of $\C$ over the $m_g$ sub-periods of $g$.}
  \label{tab:conversion}
  \begin{tabular}{lll}
    \toprule
    Rule & Weights & Typical use \\
    \midrule
    \code{sum}     & $(1,\dots,1)$            & flows (production, sales) \\
    \code{average} & $(1/m_g,\dots,1/m_g)$    & indices, rates, flows at annualised rates \\
    \code{first}   & $(1,0,\dots,0)$          & stocks measured at the start of the period \\
    \code{last}    & $(0,\dots,0,1)$          & stocks measured at the end of the period \\
    \bottomrule
  \end{tabular}
\end{table}

Every estimator in the package returns an estimate $\yhat$ that satisfies
$\C\yhat=\yl$ up to rounding error. The largest deviation
$\max_g |(\C\yhat-\yl)_g|$ is reported with every result.

\subsection{Regression-based methods}\label{sec:gls}

The regression-based methods assume
\begin{equation}\label{eq:model}
  \y = \X\bb + \bm{u}, \qquad \mathrm{E}(\bm{u})=\bm{0}, \qquad
  \Var(\bm{u}) = \sigma^2\Sig(\rho),
\end{equation}
where $\X$ ($n\times k$) contains the indicators and, by default, a constant.
Premultiplying by $\C$ gives the low-frequency regression
\begin{equation}\label{eq:lowmodel}
  \yl = \Xl\bb + \bm{u}_l, \qquad \Xl = \C\X, \qquad
  \Var(\bm{u}_l) = \sigma^2 \V, \qquad \V = \C\Sig\C\T .
\end{equation}
For a given $\rho$, the GLS estimator of $\bb$ and the best linear unbiased
predictor of $\y$ \citep{chow1971best,goldberger1962best} are
\begin{align}
  \bhat &= (\Xl\T\V^{-1}\Xl)^{-1}\Xl\T\V^{-1}\yl, \label{eq:beta}\\
  \yhat &= \X\bhat + \Sig\C\T\V^{-1}\uhat, \qquad \uhat = \yl-\Xl\bhat .\label{eq:blup}
\end{align}
In the notation of \citet{sax2013temporal}, $\yhat = \bm{p} + \bm{D}\uhat$ with
preliminary series $\bm{p}=\X\bhat$ and distribution matrix
$\bm{D}=\Sig\C\T\V^{-1}$. Since $\C\bm{D}=\I$, the estimate satisfies
\eqref{eq:constraint} exactly.

The methods differ only in the residual process, that is, in $\Sig(\rho)$
(Table~\ref{tab:structures}). Let $\Lag$ denote the $n\times n$ lag matrix
(ones on the first sub-diagonal), $\bm{H}=\I-\rho\Lag$ and $\bm{\Delta}=\I-\Lag$.
Chow--Lin assumes a stationary AR(1) process; Fernández assumes a random walk
and Litterman a random walk whose innovations follow an AR(1) process, in both
cases with zero initial conditions; OLS assumes white noise. For Chow--Lin two
normalisations of $\Sig$ are in use: the covariance form
$\rho^{|i-j|}/(1-\rho^2)$ and the correlation form $\rho^{|i-j|}$. They give
identical estimates for fixed $\rho$ and under maximum likelihood, but
different estimates of $\rho$ under the minimum residual sum of squares
criterion (Section~\ref{sec:rho}).

\begin{table}[!htbp]
  \centering\small
  \caption{Residual structures of the regression-based methods.
    $\bm{S}=\diag\big((1-\rho^2)^{-1/2},1,\dots,1\big)$ for the covariance
    form of the AR(1) process; the correlation form multiplies $\Sig$ by $1-\rho^2$.}
  \label{tab:structures}
  \begin{tabular}{llll}
    \toprule
    Method & Residual process & $\Sig(\rho)$ & Factor $\bm{T}$, $\Sig=\bm{T}\bm{T}\T$ \\
    \midrule
    OLS              & $u_t = e_t$                               & $\I$ & $\I$ \\
    Chow--Lin        & $u_t = \rho u_{t-1} + e_t$, stationary     & $[\rho^{|i-j|}/(1-\rho^2)]_{ij}$ & $\bm{H}^{-1}\bm{S}$ \\
    Fernández        & $u_t = u_{t-1} + e_t$, $u_0=0$             & $(\bm{\Delta}\T\bm{\Delta})^{-1}$ & $\bm{\Delta}^{-1}$ \\
    Litterman        & $(1-L)(1-\rho L)u_t = e_t$, zero start     & $(\bm{\Delta}\T\bm{H}\T\bm{H}\bm{\Delta})^{-1}$ & $\bm{\Delta}^{-1}\bm{H}^{-1}$ \\
    Dynamic, ADL     & see Section~\ref{sec:dynamic}              & $\bm{H}^{-1}\bm{H}^{-\top}$ & $\bm{H}^{-1}$ \\
    \bottomrule
  \end{tabular}
\end{table}

\paragraph{Computation through the triangular factor.}
Every residual structure in Table~\ref{tab:structures} has a lower-triangular
factor $\bm{T}$, $\Sig=\bm{T}\bm{T}\T$, which is a product of inverses of
bidiagonal matrices. Products $\bm{T}\bm{z}$ and $\bm{T}\T\bm{z}$ are therefore
first-order recursive filters, evaluated forwards and backwards in $O(n)$
operations. The engine computes $\bm{W}=\C\bm{T}$ ($n_l\times n$) by filtering
the rows of $\C$, obtains $\V=\bm{W}\bm{W}\T$, and writes the distribution term
in \eqref{eq:blup} as $\Sig\C\T\V^{-1}\uhat=\bm{T}\,(\bm{W}\T\V^{-1}\uhat)$.
The $n\times n$ matrix $\Sig$ is never formed, and one evaluation of the
likelihood costs $O(n\,n_l^2)$ instead of the $O(n^2 n_l)$ of the direct
formula. This makes, for example, monthly-to-daily problems with several
thousand high-frequency periods tractable (Section~\ref{sec:cost}).

\paragraph{Whitening and inference.}
With the Cholesky factorisation $\V=\bm{R}\bm{R}\T$, the transformed variables
$\tilde{\bm{y}}=\bm{R}^{-1}\yl$ and $\tilde{\X}=\bm{R}^{-1}\Xl$ satisfy an
ordinary regression with spherical errors. The coefficients are obtained by a
QR decomposition of $\tilde{\X}$ after scaling its columns to unit length, so
that rank decisions and the reported condition number do not depend on the
units of the indicators. The reported statistics follow the conventions of the
R package (Table~\ref{tab:statistics}): the weighted residual sum of squares
$\rss = \tilde{\bm{e}}\T\tilde{\bm{e}}$ with
$\tilde{\bm{e}}=\bm{R}^{-1}\uhat$, the maximum-likelihood and unbiased
estimators of $\sigma^2$, Student-$t$ tests with $n_l-k$ degrees of freedom, and
$R^2$ computed with a total sum of squares that is always centred on the
whitened constant. Two versions of the information criteria are reported: the
R definitions $\log\hat\sigma^2_{ML}+2k/n_l$ and
$\log\hat\sigma^2_{ML}+k\log(n_l)/n_l$, which depend on the scaling of $\Sig$
and are comparable only within one residual structure, and the likelihood-based
$-2\ell+2p$ and $-2\ell+p\log n_l$, which include $\log|\V|$ and are comparable
across structures. The lag-one autocorrelation of $\tilde{\bm{e}}$ is reported
as a specification check: under a correct residual model the whitened residuals
are uncorrelated.

\begin{table}[!htbp]
  \centering\small
  \caption{Statistics reported by a regression-based fit. $\tilde{\bm{e}}$ are
    the whitened residuals and $\ell$ the concentrated log-likelihood
    \eqref{eq:loglik}; $p=k+1$, plus one if $\rho$ is estimated.}
  \label{tab:statistics}
  \begin{tabularx}{\linewidth}{lX}
    \toprule
    Quantity & Definition \\
    \midrule
    Coefficient covariance & $\hat\sigma^2_{GLS}(\Xl\T\V^{-1}\Xl)^{-1}$, $\hat\sigma^2_{GLS}=\rss/(n_l-k)$ \\
    Tests and intervals & $t=\hat\beta_j/\mathrm{se}_j$ with $n_l-k$ degrees of freedom \\
    $\hat\sigma^2_{ML}$ & $\rss/n_l$ \\
    $R^2$, adjusted $R^2$ & $1-\rss/\mathrm{tss}$; $1-(1-R^2)(n_l-1)/(n_l-k)$ \\
    AIC, BIC (R definitions) & $\log\hat\sigma^2_{ML}+2k/n_l$; $\log\hat\sigma^2_{ML}+k\log(n_l)/n_l$ \\
    AIC, BIC (likelihood) & $-2\ell+2p$; $-2\ell+p\log n_l$ \\
    Residual autocorrelation & lag-one autocorrelation of $\tilde{\bm{e}}$ \\
    Design condition number & of the column-scaled whitened design \citep{belsley1980regression} \\
    Prediction standard error & $\sqrt{\diag\Var(\yhat-\y)}$, Eq.~\eqref{eq:predvar} \\
    \bottomrule
  \end{tabularx}
\end{table}

\subsection{Estimation of the autoregressive parameter}\label{sec:rho}

Three rules determine $\rho$. Under \code{maxlog}, $\rho$ maximises the
Gaussian log-likelihood of \eqref{eq:lowmodel} with $\bb$ and $\sigma^2$
concentrated out,
\begin{equation}\label{eq:loglik}
  \ell(\rho) = -\frac{n_l}{2}\left[\log(2\pi) + 1 + \log\frac{\rss(\rho)}{n_l}\right]
               - \frac{1}{2}\log|\V(\rho)|,
  \qquad \rss(\rho) = \uhat\T\V(\rho)^{-1}\uhat .
\end{equation}
Because $\sigma^2$ is concentrated out, $\ell$ is invariant to the scale of
$\yl$ and of $\Sig$, so the estimate does not depend on units or on the
normalisation of the AR(1) covariance. Under \code{minrss}, $\rho$ minimises
$\rss(\rho)$; since $\V$ is not normalised, the result depends on the scaling
of $\Sig$, and the correlation and covariance forms reproduce the variants
labelled Ecotrim and Quilis in the R package \citep{sax2013temporal}.
Under \code{fixed}, $\rho$ is supplied by the user.

The criterion is first evaluated on a grid over $[-0.999,\,0.999]$ (199 points
by default) and then refined by Brent's bounded method \citep{brent1973algorithms}
between the neighbours of the best grid point. The grid protects against
convergence to a local optimum, and the whole profile is returned with the
result. Following the R package, an estimate below a truncation value
(default $0$) is replaced by that value, since negative $\rho$ produces
alternating high-frequency residuals; the untruncated optimum is also
reported.

The estimate of $\rho$ is accompanied by identification diagnostics. Collinear
regressors and an exact fit ($\uhat=\bm{0}$, in which case $\yhat=\X\bhat$ for
every $\rho$) do not depend on $\rho$ and are detected before the search; in
the second case $\rho$ is reported as not identified. After the search, the
package flags an optimum on the boundary of the interval, a flat criterion,
and competing local optima, defined as admissible optima whose log-likelihood
lies within $\chi^2_{1,0.95}/2\approx 1.92$ of the global optimum (5\% of the
RSS under \code{minrss}), i.e.\ optima that a likelihood-ratio test would not
distinguish.

\subsection{Dynamic models}\label{sec:dynamic}

The dynamic Chow--Lin model of \citet{santossilva2001chowlin},
\begin{equation}\label{eq:dynamic}
  y_t = \rho\,y_{t-1} + \bm{x}_t\T\bb + e_t, \qquad e_t \sim \mathrm{WN}(0,\sigma^2),
\end{equation}
is solved recursively as $\y = \bm{H}^{-1}\X\bb + \bm{r}\,y_0 + \bm{H}^{-1}\bm{e}$,
with truncation remainder $r_t=\rho^t$. It is therefore a GLS problem with
filtered regressors $\bm{H}^{-1}\X$, the additional regressor $\bm{r}$ when
$\rho\neq 0$ (its coefficient is the initial value $y_0$), and
$\Sig=\bm{H}^{-1}\bm{H}^{-\top}$.

The autoregressive distributed-lag model ADL(1,1) of \citet{proietti2006temporal},
\begin{equation}\label{eq:adl}
  y_t = \rho\,y_{t-1} + \bm{x}_t\T\bb_0 + \bm{x}_{t-1}\T\bb_1 + e_t ,
\end{equation}
is handled in the same way with regressors $[\X,\ \Lag\X]$, where the lag is
applied to the non-constant columns only (the lag of a constant is collinear
with it) and $\bm{x}_0$ is set to $\bm{x}_1$. Under the common-factor
restriction $\bb_1=-\rho\bb_0$, Equation~\eqref{eq:adl} becomes
$(1-\rho L)(y_t-\bm{x}_t\T\bb_0)=e_t$, which is the Chow--Lin model with AR(1)
residuals. The restriction is tested in Section~\ref{sec:assessment}. Both
dynamic models are flagged as experimental; the ADL model is not part of the R
package and is validated against a dense reference implementation instead.

\subsection{Prediction variance}

Conditional on $\rho$, the covariance of the prediction error of
\eqref{eq:blup} is \citep{goldberger1962best}
\begin{equation}\label{eq:predvar}
  \Var(\yhat-\y) = \sigma^2\left(\Sig - \bm{G}\V^{-1}\bm{G}\T\right)
                 + \bm{Q}\,\Var(\bhat)\,\bm{Q}\T,
  \qquad \bm{G}=\Sig\C\T,\quad \bm{Q}=\X-\bm{G}\V^{-1}\Xl ,
\end{equation}
with $\Var(\bhat)=\sigma^2(\Xl\T\V^{-1}\Xl)^{-1}$ and $\sigma^2$ replaced by
$\hat\sigma^2_{GLS}$. Only the diagonal is computed, in $O(n\,n_l^2)$
operations using the factor $\bm{T}$. Pointwise prediction intervals use the
$t_{n_l-k}$ distribution. By construction $\C\Var(\yhat-\y)\C\T=\bm{0}$, the
variance is zero for the constrained sub-period under the \code{first} and
\code{last} rules, and for random-walk residuals it grows with the
extrapolation horizon. The intervals do not account for the sampling
variability of an estimated $\rho$; Section~\ref{sec:montecarlo} quantifies the
consequence.

\subsection{Benchmarking methods}\label{sec:denton}

Given one indicator $\bm{x}$ (a constant series if none is provided), the
Denton family chooses the series that preserves the movement of $\bm{x}$ while
satisfying the constraints,
\begin{equation}\label{eq:denton}
  \min_{\y}\ \|\bm{D}_h\bm{W}(\y-\bm{x})\|^2 \quad \text{subject to} \quad \C\y=\yl ,
\end{equation}
where $\bm{W}=\I$ for the additive criterion and $\bm{W}=\diag(\bm{x})^{-1}$
for the proportional criterion, and $\bm{D}_h$ is the difference operator of
order $h$ ($h=0$ levels, $h=1$ changes, $h=2$ curvature). The original
formulation of \citet{denton1971adjustment} keeps the first $h$ rows of
$\bm{D}_h$, which impose the initial condition $y_0=x_0$ and distort the start
of the series; the Denton--Cholette variant \citep{dagum2006benchmarking} drops
them and is the default. With $\bm{A}=\bm{W}\T\bm{D}_h\T\bm{D}_h\bm{W}$, the
solution solves the sparse system
\begin{equation}\label{eq:kkt}
  \begin{bmatrix} \bm{A} & \C\T \\ \C & \bm{0}\end{bmatrix}
  \begin{bmatrix} \y \\ \bm{\lambda}\end{bmatrix}
  =
  \begin{bmatrix} \bm{A}\bm{x} \\ \yl\end{bmatrix},
\end{equation}
which is solved by sparse LU decomposition without any regularisation term, so
the solution is the exact minimiser of \eqref{eq:denton}. For Denton--Cholette,
$\bm{A}$ is singular on the polynomials of degree below $h$ (scaled by
$\bm{x}$ under the proportional criterion), and \eqref{eq:kkt} is non-singular
if and only if $\C\bm{N}$ has full column rank, where $\bm{N}$ spans that null
space. The package checks this condition explicitly and reports its failure
instead of returning an arbitrary solution; the condition number of the
system is estimated with the block 1-norm estimator of
\citet{higham2000block}. Without an indicator the additive and proportional
criteria coincide, and Denton--Cholette with $h=1$ or $h=2$ is the minimum
first- or second-difference interpolation of \citet{boot1967further}, i.e.\
the smoothest series compatible with the constraints.

The uniform method distributes each value evenly,
$\yhat=\C\T(\C\C\T)^{-1}\yl$; it is defined for the \code{sum} and
\code{average} rules and does not extrapolate.

\subsection{Method specifications}

Each method is represented by an immutable specification whose fields are its
settings (Table~\ref{tab:methods}). The names of the R package are available as
presets, and every field of a preset can be overridden; a setting that does
not exist for a method raises an error listing the valid ones, and a setting
that exists but has no effect under the chosen configuration (for example a
fixed $\rho$ together with \code{maxlog} estimation) raises a warning. Every
regression method includes an intercept by default, as the R formula
\code{y \textasciitilde{} x}, and accepts any number of indicators.

\begin{table}[!htbp]
  \centering\small
  \caption{Methods and presets. Presets with suffix \code{-maxlog},
    \code{-minrss} or \code{-fixed} select the rule for $\rho$; \code{fast} is
    Chow--Lin with $\rho=0.99999$.}
  \label{tab:methods}
  \begin{tabularx}{\linewidth}{l>{\raggedright}X>{\raggedright\arraybackslash}X}
    \toprule
    Specification & Presets & Main settings (defaults) \\
    \midrule
    \code{ChowLin}        & \code{chow-lin-maxlog}, \code{-minrss-ecotrim}, \code{-minrss-quilis}, \code{-fixed}, \code{fast}
                          & intercept (yes), $\rho$ rule, fixed $\rho$ (0.5), truncation (0), covariance form \\
    \code{Litterman}      & \code{litterman-maxlog}, \code{-minrss}, \code{-fixed} & as Chow--Lin, without covariance form \\
    \code{Fernandez}      & \code{fernandez} & intercept \\
    \code{DynamicChowLin} & \code{dynamic-maxlog}, \code{-minrss}, \code{-fixed} & as Litterman, plus truncation remainder (yes); experimental \\
    \code{Proietti}       & \code{proietti-maxlog}, \code{-minrss}, \code{-fixed} & as dynamic Chow--Lin; experimental, not in R \\
    \code{OLS}            & \code{ols} & intercept \\
    \code{Denton}         & \code{denton-cholette}, \code{denton} & criterion (proportional), order $h$ (1) \\
    \code{Uniform}        & \code{uniform} & -- \\
    \bottomrule
  \end{tabularx}
\end{table}

%% file: sections/preprocessing.tex
\section{Irregular inputs, additional information and reconciliation}\label{sec:preprocessing}

\subsection{Calendar alignment and completion}

Inputs are pandas series or data frames indexed by periods (or by dates with a
known frequency), NumPy arrays with a fixed frequency ratio, or a long table
with one row per high-frequency period. The aligner maps the labels to a
calendar that records how high-frequency periods nest in low-frequency ones,
including periods of unequal length and fiscal years, and builds $\C$.
Non-numeric, infinite, duplicated or unsorted inputs are rejected.

The residual models and the difference operators assume a regular grid of
consecutive periods, and the constraint of a period requires all of its
sub-periods. By default the package therefore \emph{completes} the input:
it builds the grid from the first to the last required high-frequency period,
adds missing sub-periods of every period that has a benchmark and interior
gaps, and fills missing indicator values by linear interpolation inside the
sample and by a linear trend (estimated over the last low-frequency periods)
or a constant at the edges. A low-frequency period without a benchmark is never
given an invented one. Every added period and filled value is recorded with the
result and removed from the returned series, and completion can be disabled,
in which case the same situations raise errors.

A benchmark whose period is only partially covered by the indicators (for
example, a first year whose indicator starts in April) is ambiguous: the value
may refer to the whole period or to the observed sub-periods only. The user
chooses the interpretation explicitly: the constraint applies to the whole
period, whose missing sub-periods are added (default); the constraint applies
to the observed sub-periods with the conversion rule restricted to them; or the
input is rejected.

\subsection{Padding}\label{sec:padding}

Optionally, low-frequency periods can be added before the first and after the
last benchmark, with indicators extended and artificial benchmarks. Artificial
benchmarks are necessary for the padding to have any effect: for Chow--Lin, for
Litterman and Fernández with an intercept, and for Denton--Cholette, adding
unconstrained periods outside the sample leaves the in-sample estimate
unchanged, because neither the in-sample block of $\Sig$ nor the constrained
optimum depends on them (for random-walk models the shifted initial condition
is absorbed by the intercept). The test suite verifies this invariance
numerically.

The artificial benchmark of a padded period is a linear trend of the last
benchmarks, the edge benchmark repeated, or the benchmark-to-indicator (BI)
ratio rule recommended by the IMF Quarterly National Accounts Manual
\citep[ch.~6]{bloem2001qna}: with $r_g = y_{l,g}/(\C\bm{x})_g$ the BI ratio of
each observed period, the padded benchmark is $\hat r\,(\C\bm{x})_{g'}$, where
$\hat r$ is the ratio of the edge period or a linear projection of the last
ratios. The ratio rule lets the indicator carry the movement and projects only
the ratio, which is usually more stable than the level; projecting a trend in
the ratio is riskier, as the manual warns. Padding is a heuristic rather than
an estimator. Regression statistics computed with padded periods count the
artificial benchmarks as observations; their number is reported, and a warning
is issued.

\subsection{Additional exact constraints and aggregation weights}

Information beyond the benchmarks---known high-frequency values, benchmarks at
an intermediate frequency (for example, some quarterly totals when the target
is monthly and the benchmarks are annual), or any linear relation
$\bm{a}\T\y=b$---is appended as additional rows of $\C$ and $\yl$. All
estimators use the stacked system directly: the GLS formulas
\eqref{eq:beta}--\eqref{eq:predvar} and the likelihood keep their form, and the
Denton system \eqref{eq:kkt} gains rows. Imposing the information during
estimation keeps the estimator optimal, which estimating first and adjusting
afterwards does not. Before estimation the stacked system is processed row by
row with modified Gram--Schmidt, benchmarks first. A row that is a linear
combination of earlier rows is redundant if its value is the implied one, and
is then dropped with a warning (redundant rows make $\V$ singular); otherwise
the system is contradictory, and the error reports the conflicting values.

Aggregation weights $w_t$ generalise the conversion rules: a \code{sum}
benchmark becomes $\sum_t w_t y_t$ and an \code{average} benchmark
$\sum_t w_t y_t/\sum_t w_t$. Their main use is for rates. If an annual rate is
defined as a ratio of annual totals, for example an unemployment rate
$r_a = \sum_m N_m r_m / \sum_m N_m$ with the labour force $N_m$ as denominator,
it is a weighted average of the monthly rates and is disaggregated with the
denominator as weight. When both annual totals are known, the numerator and
the denominator can instead be disaggregated separately, each with its own
model, and divided; with the \code{sum} rule the ratio of the annual sums of
the two estimates reproduces the official annual ratio exactly.

Structural breaks are represented in regression methods by step or impulse
dummies added to the indicators. Their coefficients are estimated from the
low-frequency data, so a break must be supported by benchmarks on both sides.

\subsection{Non-negative reconciliation}

The estimators in Section~\ref{sec:framework} do not impose sign restrictions.
For series that cannot be negative, the package computes the exact projection
\begin{equation}\label{eq:nonneg}
  \min_{\y}\ \|\y-\yhat\|^2 \quad \text{subject to} \quad \C\y=\yl,\quad \y\geq\bm{0}.
\end{equation}
When every high-frequency period belongs to at most one constraint row, the
problem separates by period. For a group with positive weights $c_i$ the
Karush--Kuhn--Tucker conditions give $y_i=\max(0,\hat y_i+\lambda c_i)$, and
$\sum_i c_i y_i(\lambda)$ is non-decreasing in $\lambda$; the multiplier is
bracketed by bisection and computed in closed form on the identified active
set. For overlapping constraints (for example, known quarters inside years),
Dykstra's alternating projections \citep{dykstra1983algorithm} between the
affine set and the non-negative orthant converge to the projection onto their
intersection; the iterate is then refined in closed form on its active set and
accepted only if the KKT conditions hold. A negative benchmark, or a set of
constraints without a non-negative solution, raises an error instead of
returning an approximation.

\subsection{Cross-sectional reconciliation}

Series disaggregated separately satisfy their own constraints $\C_s\y_s=\bm{a}_s$
but not, in general, a known high-frequency total $\sum_s\y_s=\bm{z}$.
Following the two-step approach of \citet{difonzo2011simultaneous}, the
reconciled system solves
\begin{equation}\label{eq:crosssection}
  \min_{\y_1,\dots,\y_S}\ \sum_{s}\|\bm{D}\bm{W}_s(\y_s-\bm{p}_s)\|^2
  \quad\text{subject to}\quad \C_s\y_s=\bm{a}_s\ (s=1,\dots,S),\quad \sum_s\y_s=\bm{z},
\end{equation}
where $\bm{p}_s$ are the first-step estimates, $\bm{W}_s=\I$ (additive) or
$\diag(\bm{p}_s)^{-1}$ (proportional), and $\bm{D}$ is the identity (minimal
adjustments) or the first difference without initial condition (movement
preservation). The constraints overlap: in every low-frequency period where all
series have benchmarks, $\sum_s a_{s,g}$ must equal $(\C\bm{z})_g$. A mismatch is
reported period by period; otherwise one redundant temporal row per such
period is removed and the sparse KKT system is solved by LU decomposition.

\subsection{Splicing}

Disaggregation is frequently preceded by the linking of a series across a
methodological break. The package provides level, ratio and affine linking
over an overlap period, and the interpolated method of \citet{delafuente2014mixed},
in which the discrepancy between the legacy and the current series is assumed
to accumulate linearly from an anchor period at which the legacy series is
taken as correct. Observed values of the current series are never modified.

%% file: sections/assessment.tex
\section{Model assessment}\label{sec:assessment}

Because every estimator satisfies \eqref{eq:constraint} exactly, the in-sample
error of the aggregates is zero by construction and cannot be used to compare
methods, nor to weight them in a combination. The high-frequency path itself is
not observed. The package therefore provides tools that use information the
fit has not seen, or that describe properties of the estimate that matter in
production \citep{quilis2018temporal}.

\paragraph{Backtesting.}
For each of the last $H$ observed low-frequency periods, the candidate model is
re-estimated with the benchmarks up to the previous period only, the
high-frequency estimate of the held-out period (an extrapolation) is
aggregated with $\C$, and the result is compared with the actual benchmark.
Bias, mean absolute error, root mean squared error and mean absolute percentage
error are reported per candidate. Each candidate is a model together with its
own indicators, so that indicator sets can be compared as well as methods. The
backtest measures how well a model extrapolates the aggregate; it says nothing
about the distribution within periods, which is unobservable without a
high-frequency benchmark.

\paragraph{Revision analysis.}
A published high-frequency series is revised whenever a new benchmark arrives.
The revision analysis replays this process: for $v=0,\dots,V$ each candidate is
fitted with the benchmarks up to period $T-V+v$, and consecutive vintages are
compared on their common periods. Revisions are measured in levels, as a
percentage of the previous vintage, and in period-on-period changes, in
percentage points of growth for positive series and in differences otherwise.
The summary reports mean absolute revisions over all common periods and over
the last benchmarked period of the earlier vintage, i.e.\ the most recent
estimates published before the new benchmark arrived, which are the ones most
affected in practice.
Periods before the first benchmark are excluded by default, since their
revisions reflect the edge of the sample rather than the new information.
Optionally each vintage includes the indicator for the next period, so that
the revision of the first (extrapolated) estimate is measured.

\paragraph{Conformity.}
For each indicator the package compares the estimate with the indicator through
the correlation of levels, the correlation of period-on-period changes, the
standard deviations of the changes of both series and their ratio (the
volatility ratio). A disaggregation that conforms to its indicator tracks its
short-term movement without adding volatility that the indicator does not have.

\paragraph{Common-factor test.}
In the ADL(1,1) model \eqref{eq:adl}, Chow--Lin corresponds to the $q$
restrictions $\bm{g}(\bb)=\bb_1+\rho\bb_0=\bm{0}$, with $q$ the number of
lagged indicators. Conditional on $\hat\rho$, the Wald statistic
\begin{equation}\label{eq:wald}
  W = \bm{g}(\bhat)\T\left[\bm{J}\,\widehat{\Var}(\bhat)\,\bm{J}\T\right]^{-1}\bm{g}(\bhat),
  \qquad \bm{J} = \begin{bmatrix}\hat\rho\,\I_q & \I_q\end{bmatrix},
\end{equation}
where $\bm{J}$ is the Jacobian of $\bm{g}$ with respect to $(\bb_0,\bb_1)$, is
reported as $W/q$ with reference distribution $F(q,\,n_l-k)$. Rejection
indicates that the dynamics of the indicators are not transmitted one to one to
the target. It is not, by itself, a reason to prefer the ADL estimate, whose
volatility can exceed that of the indicator (Section~\ref{sec:diagnostics}).

\paragraph{Profile of the autoregressive parameter.}
The profile of the criterion in $\rho$, returned with every fit and plotted on
request, shows whether $\rho$ is identified. \citet{quilis2018temporal} notes
that the intercept and a persistent AR(1) process compete to explain the level
of the target, so that removing the intercept can sharpen the likelihood; the
example in Section~\ref{sec:diagnostics} illustrates this.

%% file: sections/software.tex
\section{Software design}\label{sec:software}

\subsection{Architecture}

The package separates data handling, estimation and reporting
(Figure~\ref{fig:architecture}). An aligner validates the inputs and, together
with a grid builder, produces an immutable data object holding $\yl$, the
high-frequency regressors, $\C$, the type of each constraint row and the record
of every artificial value. A method specification
(Table~\ref{tab:methods}) builds one of three estimators: the GLS engine, which
receives a residual structure and a rule for $\rho$; the Denton solver; or the
uniform distributor. The estimator returns arrays and a log of diagnostics, from
which immutable result objects are assembled. Reconciliation and assessment
operate on results.

\begin{figure}[!htbp]
  \centering
  \begin{tikzpicture}[
      font=\small,
      box/.style={draw=black!55, rounded corners=2pt, fill=#1, align=center,
                  minimum height=1.05cm, text width=2.55cm, inner sep=3pt},
      box/.default={black!3},
      arr/.style={-{Stealth[length=2.2mm]}, thick, black!65},
      node distance=0.45cm and 0.55cm]
    \node[box] (in) {Inputs\\ \footnotesize pandas, NumPy, long table};
    \node[box, right=of in] (align) {Aligner and grid\\ \footnotesize calendar, completion, padding, constraints};
    \node[box, right=of align] (data) {Data\\ \footnotesize $\yl$, $\X$, $\C$, record of artificial values};
    \node[box, right=of data] (est) {Estimator\\ \footnotesize GLS engine, Denton, uniform};
    \node[box, below=0.9cm of est] (res) {Result\\ \footnotesize estimate, inference, $\rho$ profile, diagnostics};
    \node[box, left=of res] (spec) {Method specification\\ \footnotesize presets, settings, validation};
    \node[box, left=of spec] (assess) {Assessment\\ \footnotesize backtest, revisions, conformity, tests};
    \node[box, left=of assess] (rec) {Reconciliation\\ \footnotesize non-negative, cross-sectional};
    \draw[arr] (in) -- (align);
    \draw[arr] (align) -- (data);
    \draw[arr] (data) -- (est);
    \draw[arr] (spec.north) |- ++(0,0.45) -| ([xshift=-6mm]est.south);
    \draw[arr] ([xshift=6mm]est.south) -- ([xshift=6mm]res.north);
    \draw[arr] (res.south) -- ++(0,-0.35) -| (assess.south);
    \draw[arr] (res.south) -- ++(0,-0.35) -| (rec.south);
  \end{tikzpicture}
  \caption{Data flow. Every object passed between stages is immutable; warnings
    raised by any stage are stored in the result.}
  \label{fig:architecture}
\end{figure}

The implementation follows a small set of rules. There is one engine per
family: GLS methods differ only in a residual structure and a rule for $\rho$,
so the GLS algebra exists once, and a new method is added by writing a
structure that returns its triangular factor. Behaviour is explicit: inputs are
not imputed, no fallback method is used, and no pseudo-inverse is computed.
Failures raise typed exceptions (for example, a singular design, an infeasible
system of constraints or an unidentified Denton problem) with an explanation;
statistical problems raise typed warnings that are also stored in the result
(Table~\ref{tab:warnings}). Specifications and results are frozen dataclasses,
composed of smaller statistics objects rather than organised by inheritance.
Every optimised computation has a dense textbook counterpart in the test suite.
The package is fully type-annotated and its only run-time dependencies are
NumPy, SciPy and pandas.

\begin{table}[!htbp]
  \centering\small
  \caption{Statistical warnings. Each is raised when it occurs and stored with the result.}
  \label{tab:warnings}
  \begin{tabularx}{\linewidth}{lX}
    \toprule
    Warning & Condition \\
    \midrule
    \code{RhoBoundaryWarning} & $\hat\rho$ on the boundary of the search interval, often a sign of a misspecified residual process \\
    \code{RhoNotIdentifiedWarning} & flat criterion in $\rho$ \\
    \code{MultipleOptimaWarning} & admissible local optima not distinguishable by a likelihood-ratio test \\
    \code{ExactFitWarning} & zero low-frequency residuals; $\rho$ not identified \\
    \code{IllConditionedWarning} & nearly collinear regressors or nearly singular $\V$ or KKT system \\
    \code{IndicatorSignWarning} & proportional Denton with an indicator that changes sign \\
    \code{ArtificialDataWarning} & values added by completion or padding \\
    \code{MissingTargetWarning} & interior low-frequency periods without benchmark \\
    \code{RedundantConstraintWarning} & consistent constraint implied by others, dropped \\
    \code{IneffectiveParameterWarning} & setting without effect under the chosen configuration \\
    \code{ExperimentalMethodWarning} & dynamic and ADL methods \\
    \bottomrule
  \end{tabularx}
\end{table}

\subsection{Interface}

The main entry point is \code{TemporalDisaggregator}, which takes a preset name
or a specification, the conversion rule and optional overrides, and whose
\code{fit} method returns a \code{DisaggregationResult}.
Listing~\ref{lst:basic} disaggregates the annual sales of the Swiss
pharmaceutical industry with quarterly exports, the example of the R package
\citep{sax2013temporal}; Listing~\ref{lst:summary} shows the printed summary.
For automated pipelines, the function \code{disaggregate} returns only the
series.

\begin{lstlisting}[caption={Basic use.},label={lst:basic}]
import tempdisagg as td
from tempdisagg.datasets import load_swisspharma

sales, trade = load_swisspharma()      # annual Series, quarterly DataFrame
result = td.TemporalDisaggregator("chow-lin-maxlog").fit(sales, trade[["exports"]])

result.values                          # quarterly estimate, pd.Series with PeriodIndex
result.prediction_interval(0.95)       # estimate with lower and upper bounds
result.regression.table()              # coefficients with inference
result.rho.profile                     # criterion on the grid of rho
result.to_frame()                      # flags extrapolated and filled periods

series = td.disaggregate(sales, trade[["exports"]], method="chow-lin-maxlog")
\end{lstlisting}

\begin{lstlisting}[style=output,float=htbp,caption={Output of \code{print(result.summary())} for Listing~\ref{lst:basic}.},label={lst:summary}]
'chow-lin-maxlog' disaggregation with 'sum' conversion
36 low-freq. obs. converted to 158 high-freq. obs.

Residuals (low frequency):
    Min      1Q  Median     3Q    Max
-77.892 -7.7108 -4.6276 9.6465 36.448

Coefficients:
             Estimate  Std. Error  t value    Pr(>|t|)
(Intercept)   12.4089     1.49303  8.31119 1.06081e-09  ***
exports     0.0133918 0.000167167  80.1106 2.53953e-40  ***
---
Signif. codes: 0 '***' 0.001 '**' 0.01 '*' 0.05 '.' 0.1 ' ' 1

Residual std. error (GLS): 10.4417 on 34 degrees of freedom
Multiple R-squared: 0.9947   Adjusted R-squared: 0.9946
Log-likelihood: -159.4555   AIC: 4.7456   BIC: 4.8335   (likelihood-based AIC: 326.9109)
Lag-1 autocorrelation of whitened residuals: -0.0804
AR1-Parameter: 0.000000 (estimated by maxlog, truncated (unconstrained -0.306953))
Max. aggregation error: 3.411e-13
\end{lstlisting}

Listing~\ref{lst:config} shows how settings are inspected and overridden.
Listing~\ref{lst:constraints} adds a known quarterly value, reconciles to
non-negativity and configures padding with the BI-ratio rule.
Listing~\ref{lst:assessment} compares two candidates with the assessment tools
of Section~\ref{sec:assessment}.

\begin{lstlisting}[caption={Method specifications and settings.},label={lst:config}]
from tempdisagg.methods import ChowLin, make_method

ChowLin().include_intercept                          # True
make_method("fernandez", include_intercept=False)    # Fernandez(include_intercept=False)
td.TemporalDisaggregator("chow-lin-maxlog", truncated_rho=-1)          # keep negative rho
td.TemporalDisaggregator("denton-cholette", criterion="additive", h=2)
td.TemporalDisaggregator("chow-lin-fixed", fixed_rho=0.5).describe()   # settings and R names
\end{lstlisting}

\begin{lstlisting}[caption={Additional constraints, non-negativity and padding.},label={lst:constraints}]
import pandas as pd

known = td.known_values(pd.Series([30.0], index=pd.PeriodIndex(["1975Q1"], freq="Q")))
informed = td.TemporalDisaggregator().fit(sales, trade[["exports"]], constraints=known)

non_negative = td.NonNegativeReconciler().reconcile(result)

padding = td.PaddingSettings(before=0, after=1, target_method="ratio")
padded = td.TemporalDisaggregator("denton-cholette", padding=padding)
\end{lstlisting}

\begin{lstlisting}[caption={Assessment tools.},label={lst:assessment}]
candidates = {
    "chow-lin": td.Candidate(td.TemporalDisaggregator("chow-lin-maxlog"), trade[["exports"]]),
    "denton": td.Candidate(td.TemporalDisaggregator("denton-cholette"), trade[["exports"]]),
}
td.backtest(sales, candidates, holdout=5).summary()
td.revision_analysis(sales, candidates, n_vintages=8).summary()
td.conformity(result)

adl = td.TemporalDisaggregator("proietti-maxlog").fit(sales, trade[["exports"]])
print(td.common_factor_test(adl).summary())
\end{lstlisting}

\subsection{Testing}\label{sec:testing}

The test suite (420 test cases) is organised in layers.
\emph{Agreement with R}: the coefficients, standard errors, $t$ statistics,
adjusted $R^2$ and residual quantiles printed in the vignette of the R package
are asserted directly, and 29 configurations covering every shared method, the
three rules for $\rho$ and the four conversion rules are compared
with reference outputs generated by a script with R~4.5.3 and \pkg{tempdisagg}~1.2.0
(Section~\ref{sec:parity}). A separate continuous-integration workflow installs
R, regenerates the references and runs the comparison.
\emph{Reference engine}: every optimised computation (estimates, coefficient
covariance, prediction variance, log-likelihood, optima in $\rho$, Denton
solutions, ADL design) is compared with a dense implementation of the textbook
formulas. \emph{Statistical properties}: Monte Carlo experiments check the
coverage of prediction and coefficient intervals, the consistency of the
maximum-likelihood estimate of $\rho$, and invariance to units, column order
and constants absorbed by the intercept. \emph{Robustness}: zero, constant and
collinear indicators, exact fits, extreme scales, boundary values of $\rho$,
malformed inputs and unidentified Denton problems are tested, and a randomised
contract test draws 300 problems with random sizes, frequency ratios, scales
from $10^{-6}$ to $10^{6}$, conversion rules, methods, duplicated or null
indicator columns, missing benchmarks and indicator values, padding and known
values; each fit must either return finite values that satisfy the constraints
or raise one of the package's exceptions. The suite, a linter and a static type
checker run in continuous integration on Python 3.10 to 3.13.

%% file: sections/validation.tex
\section{Validation}\label{sec:validation}

This section reports five checks: agreement with the R package, recovery of
series whose high-frequency values are known, Monte Carlo properties of the
inference, the assessment tools on the R package's example data, and
computational cost. All results were produced with version 0.6.6 by the
ancillary script \code{anc/reproduce\_paper.py}; Monte Carlo experiments use
500 replications with fixed seeds.

\subsection{Agreement with the R package}\label{sec:parity}

\begin{figure}[!htbp]
  \centering
  \includegraphics[width=0.82\linewidth]{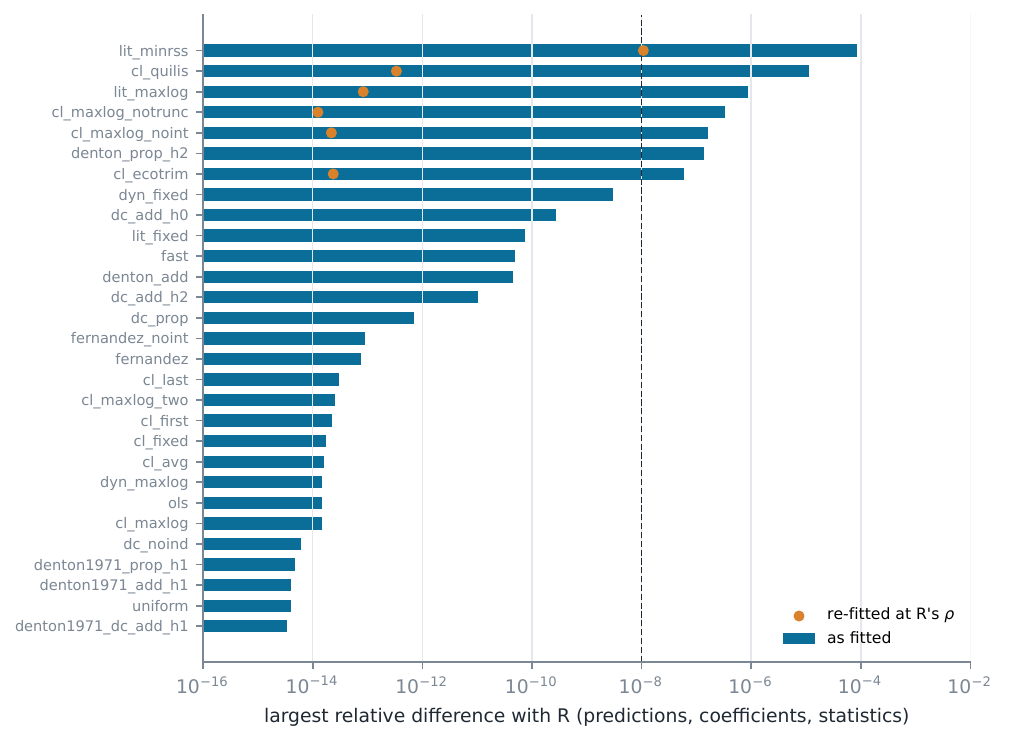}
  \caption{Largest relative difference with R per case (log scale). The dashed
    line marks the tolerance of the test suite, $10^{-8}$.}
  \label{fig:parity}
\end{figure}

The reference outputs cover 29 configurations: 26 on the Swiss pharmaceutical
data distributed with the R package (annual sales 1975--2010 and quarterly
exports and imports 1972Q1--2011Q2), which include retropolated and
extrapolated quarters, and three on the numerical example of
\citet{denton1971adjustment}. For each case the comparison covers the
predictions over the full output and, for regression methods, the
coefficients, standard errors, $\hat\sigma^2_{ML}$, $\hat\sigma^2_{GLS}$, RSS,
TSS, $R^2$, adjusted $R^2$, log-likelihood, AIC and BIC. The largest relative
difference over all these quantities is reported in Table~\ref{tab:parity} and
Figure~\ref{fig:parity} (absolute differences for $R^2$, which can be close to
zero).

With $\rho$ fixed or absent (23 cases, including those in which the estimate
is truncated at zero and therefore coincides in both implementations), the
largest relative difference ranges from $3\times10^{-15}$ to $3\times10^{-9}$,
with one exception: the original proportional Denton method with $h=2$
($1.4\times10^{-7}$), where the R solution satisfies its own aggregation
constraint only to about $2\times10^{-5}$ because of an ill-conditioned dense
solve, whereas the sparse solution of Section~\ref{sec:denton} satisfies it to
$10^{-13}$. With $\rho$ estimated and not truncated (six cases), the estimates
differ from those of R by up to $4\times10^{-6}$, and the outputs by up to
$9\times10^{-5}$. These differences are caused by the stopping tolerance of the
one-dimensional optimiser used by R, which is of the order of $10^{-4}$; the test
suite verifies that the criterion at the Python optimum is never worse than at
R's. Re-fitting the same specification at R's value of $\rho$ reduces the
differences to between $10^{-14}$ and $4\times10^{-13}$, except for Litterman
with the minimum-RSS criterion ($1.1\times10^{-8}$), whose slope is sensitive
to the digits of $\rho$ printed by R.

\begin{table}[!htbp]
  \centering
  \caption{Agreement with R~4.5.3 / \pkg{tempdisagg}~1.2.0. ``As fitted'' is the
    largest relative difference over predictions, coefficients, standard errors
    and reported statistics; ``At R's $\rho$'' repeats the comparison after
    re-fitting at R's estimate of $\rho$ (estimated, untruncated cases only).
    Settings not listed take their default values. Swiss cases use exports as
    indicator, \code{cl\_maxlog\_two} also imports, and \code{dc\_noind} and
    \code{uniform} no indicator; cases \code{denton1971\_*} use the example of
    \citet{denton1971adjustment}.}
  \label{tab:parity}
  \resizebox{\linewidth}{!}{\input{tables/r_parity}}
\end{table}

\subsection{Recovery of known high-frequency series}\label{sec:recovery}

Since the high-frequency path of a disaggregated series is unobservable, two
series that are observed quarterly were aggregated to annual frequency and
disaggregated back, which allows direct measurement of the error
(Tables~\ref{tab:recovery_exports} and~\ref{tab:recovery_gdp}). The
experiments are illustrations on two series, not a general ranking of methods.

\paragraph{Swiss exports from imports.}
Quarterly exports of the Swiss pharmaceutical industry (1975--2010) were summed
to annual values and disaggregated with quarterly imports as the indicator.
Uniform distribution, which ignores the indicator, gives the smallest errors
(level RMSE 361.9 and growth RMSE 5.8 percentage points, against 429.7 and 6.8
for Chow--Lin). The quarterly growth rates of imports are only weakly related to
those of exports, so following the indicator adds noise. The example shows why
the choice of indicator dominates the choice of method, and why tools that
assess the indicator (Section~\ref{sec:diagnostics}) are needed.

\begin{figure}[!htbp]
  \centering
  \includegraphics[width=0.9\linewidth]{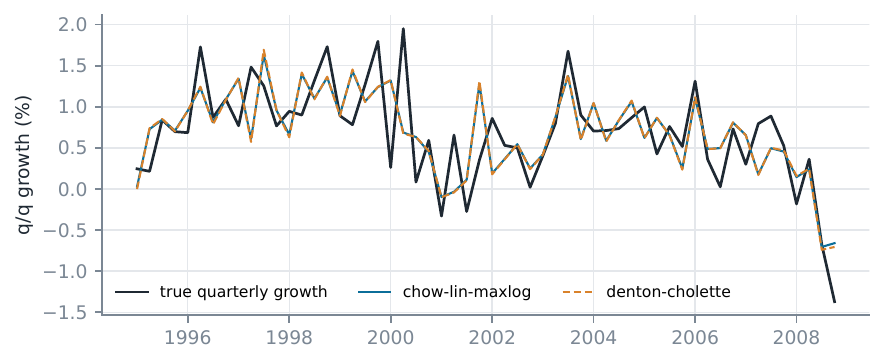}
  \caption{Quarter-on-quarter growth of U.S.\ real GDP, 1995--2008: true values
    and the estimates of Chow--Lin and Denton--Cholette from annual averages.}
  \label{fig:gdp}
\end{figure}

\paragraph{U.S.\ real GDP from consumption.}
Quarterly U.S.\ real GDP from the \pkg{statsmodels} macroeconomic data set
\citep{seabold2010statsmodels} was averaged to annual values for 1959--2008 (50
years, 200 quarters) and disaggregated with real personal consumption as
indicator under the \code{average} rule; the indicator for 2009Q1--Q3 was used
to extrapolate the three quarters following the last benchmark. Here the
indicator is informative: Chow--Lin, Fernández, Litterman, the ADL model and
Denton--Cholette reproduce quarterly growth with an RMSE of about 0.6
percentage points and a correlation of about 0.74 with the true growth rates,
against 1.7 points and 0.31 for uniform distribution (Figure~\ref{fig:gdp}).
The level error of the original Denton method is about 1.7 times that of
Denton--Cholette (47.1 against 27.4) because of its initial condition. The dynamic Chow--Lin
model is the least accurate of the indicator-based methods, in sample and in
extrapolation. The extrapolation errors for 2009, a recession in which GDP fell
more than consumption, are of the order of 300 (about 2\% of the level) for all
methods except the dynamic model (410).

\begin{table}[!htbp]
  \centering\small
  \caption{Quarterly Swiss exports (1975--2010) recovered from annual totals
    with imports as indicator. Growth refers to quarter-on-quarter percentage
    changes; $\hat\rho$ is the value used after truncation.}
  \label{tab:recovery_exports}
  \input{tables/recovery_exports}
\end{table}

\begin{table}[!htbp]
  \centering\small
  \caption{Quarterly U.S.\ real GDP (1959--2008, billions of chained 2005
    dollars) recovered from annual averages with real consumption as indicator;
    the last column is the RMSE of the extrapolation of 2009Q1--Q3.}
  \label{tab:recovery_gdp}
  \resizebox{\linewidth}{!}{\input{tables/recovery_gdp}}
\end{table}

\subsection{Monte Carlo properties}\label{sec:montecarlo}

Data were simulated from the Chow--Lin model with $\rho=0.7$, an intercept of 5,
a slope of 1.5 on a random-walk indicator, 30 annual and 120 quarterly
observations. Table~\ref{tab:coverage} reports the empirical coverage of
nominal 95\% intervals. With $\rho$ known, the prediction intervals of
\eqref{eq:predvar} have coverage 95.2\% and the slope intervals
93.8\%. With $\rho$ estimated by maximum likelihood, coverage is
95.8\% and 89.8\%, respectively. With 500 replications the Monte Carlo standard error of a
coverage of 95\% is about one percentage point for the slope; the prediction
coverage pools 120 correlated quarters per replication. Conditioning on an
estimated $\rho$ therefore leaves the prediction intervals close to the
nominal level in this design, but the slope intervals under-cover by about five
points, because they ignore the uncertainty of $\hat\rho$.

\begin{table}[!htbp]
  \centering\small
  \caption{Empirical coverage of nominal 95\% intervals for the high-frequency
    values and for the slope (Chow--Lin data, $\rho=0.7$, $n_l=30$).}
  \label{tab:coverage}
  \input{tables/mc_coverage}
\end{table}

For the common-factor test (Table~\ref{tab:commonfactor}), the size was
estimated with data from the Chow--Lin model and the power with data from the
ADL(1,1) model with $\rho=0.6$, $\beta_0=1$ and $\beta_1=0.4$, for which the
restriction fails. The test is slightly liberal in small samples (rejection
rates of 7.0\%, 6.6\% and 5.6\% at a nominal 5\% for $n_l=20$, 40 and 80),
which is consistent with conditioning on $\hat\rho$, and rejects the false
restriction in every replication.

\begin{table}[!htbp]
  \centering\small
  \caption{Rejection frequencies of the common-factor test at the 5\% level.}
  \label{tab:commonfactor}
  \input{tables/mc_common_factor}
\end{table}

\subsection{Assessment tools on the example data}\label{sec:diagnostics}

\begin{figure}[!htbp]
  \centering
  \includegraphics[width=0.95\linewidth]{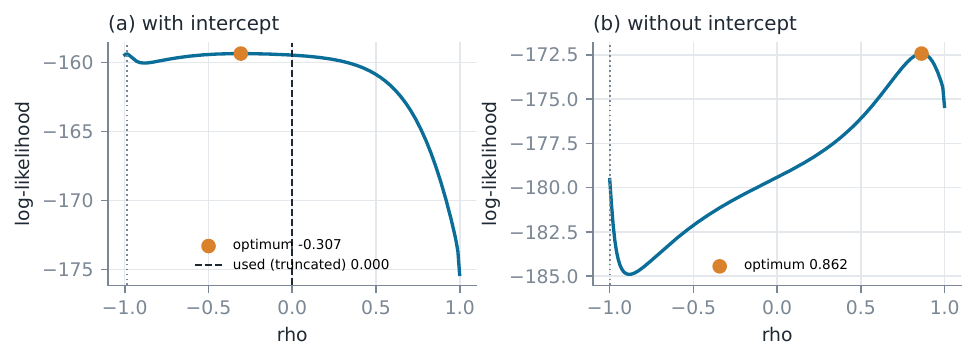}
  \caption{Profile of the log-likelihood in $\rho$ for Chow--Lin on the Swiss
    data (sales on exports), with and without intercept.}
  \label{fig:rhoprofile}
\end{figure}

\paragraph{Estimated parameter and intercept.}
For the Chow--Lin model of Listing~\ref{lst:basic}, the maximum-likelihood
estimate of $\rho$ is $-0.307$, which is truncated to zero, and the profile of
the log-likelihood is nearly flat over most of the interval
(Figure~\ref{fig:rhoprofile}a). Without an intercept the profile has a clear
maximum at $\hat\rho=0.862$ (Figure~\ref{fig:rhoprofile}b), consistent with
the observation of \citet{quilis2018temporal} that the constant and a
persistent AR(1) process compete to explain the level. Removing the intercept
is appropriate only when the indicator carries the level of the target.

\paragraph{Backtest.}
With five rolling origins (Table~\ref{tab:backtest}), Chow--Lin with exports has
the smallest RMSE of the annual extrapolation (63.7); adding imports increases
it to 70.8, in line with the weak relation between imports and the target found
above, and Fernández and Denton--Cholette with exports give 78.8 and 82.6.

\begin{table}[!htbp]
  \centering\small
  \caption{Rolling one-step-ahead backtest of the annual sales (last five years held out in turn).}
  \label{tab:backtest}
  \input{tables/swiss_backtest}
\end{table}

\paragraph{Conformity and revisions.}
Table~\ref{tab:conformity} compares the quarterly estimates with exports.
Denton--Cholette reproduces the volatility of the growth rate of exports
(ratio 1.00), Chow--Lin attenuates it (0.87), Fernández attenuates it further
(0.61), as does Litterman, whose estimate of $\rho$ is truncated to zero so that
it coincides with Fernández, and the ADL
model amplifies it (1.28) because it also uses the lagged indicator. The dynamic
model coincides with Chow--Lin here because its estimate of $\rho$ is truncated
at zero. For the same models, Table~\ref{tab:revisions} replays the arrival of
the last eight benchmarks. Over all common quarters Denton--Cholette revises
least, since a new benchmark affects mainly the adjacent quarters; for the
quarters of the last year benchmarked before each new figure, Chow--Lin
revises least. The dynamic model has the
largest revisions on every measure, which agrees with the ranking reported by
\citet{quilis2018temporal} for Spanish data. On the same data the common-factor
test does not reject the Chow--Lin restriction ($F(1,33)=1.00$, $p=0.32$).

\begin{table}[!htbp]
  \centering\small
  \caption{Conformity of the quarterly estimates of sales with exports (growth in \%).}
  \label{tab:conformity}
  \input{tables/swiss_conformity}
\end{table}

\begin{table}[!htbp]
  \centering\small
  \caption{Mean absolute revisions (MAR) over eight vintages: levels in \% of
    the previous vintage, growth in percentage points; ``latest'' refers to the
    quarters of the last benchmarked year of the earlier vintage.}
  \label{tab:revisions}
  \resizebox{\linewidth}{!}{\input{tables/swiss_revisions}}
\end{table}

\begin{figure}[!htbp]
  \centering
  \includegraphics[width=0.95\linewidth]{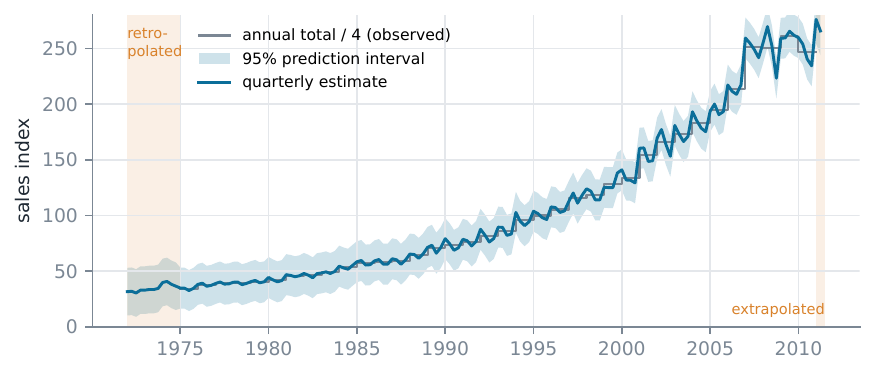}
  \caption{Quarterly sales estimated by Chow--Lin from annual sales and
    quarterly exports, with 95\% prediction intervals (conditional on $\rho$).
    Shaded areas mark quarters outside the benchmarked years.}
  \label{fig:swiss}
\end{figure}

Figure~\ref{fig:swiss} shows the corresponding estimate and its prediction
intervals. The intervals are widest in the retropolated and extrapolated
quarters, where no benchmark constrains the estimate.

\subsection{Computational cost}\label{sec:cost}

Table~\ref{tab:timing} reports the median time of five fits for simulated
problems of increasing size. Denton--Cholette, a single sparse solve, takes at
most 0.01 seconds in every case. Chow--Lin with maximum-likelihood estimation
of $\rho$ evaluates the likelihood at the 199 grid points and during the
refinement; it takes between 0.1 and 0.25 seconds for annual, quarterly and
monthly problems, and 2.8 seconds for a monthly-to-daily problem with 3{,}600
high-frequency periods, for which forming $\Sig$ explicitly would require a
$3600\times3600$ matrix per evaluation.

\begin{table}[!htbp]
  \centering\small
  \caption{Median time of one fit, including the search over $\rho$ (199 grid
    points and Brent refinement) for Chow--Lin. Simulated data; Intel Xeon at 2.1\,GHz (two cores),
Python 3.13, NumPy 2.5, SciPy 1.18.}
  \label{tab:timing}
  \input{tables/timing}
\end{table}

%% file: tables/r_parity.tex
\begin{tabular}{lllcrrrr}
\toprule
Case & Preset & Settings & Conv. & $\hat\rho$ (Py) & $\hat\rho$ (R) & As fitted & At R's $\rho$ \\
\midrule
\texttt{cl\_avg} & \texttt{chow-lin-maxlog} & -- & average & 0.0000 & 0.0000 & $1.6\times 10^{-14}$ & -- \\
\texttt{cl\_ecotrim} & \texttt{chow-lin-minrss-ecotrim} & truncated\_rho=-1 & sum & 0.6043 & 0.6043 & $5.9\times 10^{-8}$ & $2.4\times 10^{-14}$ \\
\texttt{cl\_first} & \texttt{chow-lin-fixed} & fixed\_rho=0.6 & first & 0.6000 & 0.6000 & $2.2\times 10^{-14}$ & -- \\
\texttt{cl\_fixed} & \texttt{chow-lin-fixed} & fixed\_rho=0.5 & sum & 0.5000 & 0.5000 & $1.7\times 10^{-14}$ & -- \\
\texttt{cl\_last} & \texttt{chow-lin-fixed} & fixed\_rho=0.6 & last & 0.6000 & 0.6000 & $3.0\times 10^{-14}$ & -- \\
\texttt{cl\_maxlog} & \texttt{chow-lin-maxlog} & -- & sum & 0.0000 & 0.0000 & $1.5\times 10^{-14}$ & -- \\
\texttt{cl\_maxlog\_noint} & \texttt{chow-lin-maxlog} & include\_intercept=False & sum & 0.8620 & 0.8620 & $1.7\times 10^{-7}$ & $2.2\times 10^{-14}$ \\
\texttt{cl\_maxlog\_notrunc} & \texttt{chow-lin-maxlog} & truncated\_rho=-1 & sum & -0.3070 & -0.3070 & $3.4\times 10^{-7}$ & $1.2\times 10^{-14}$ \\
\texttt{cl\_maxlog\_two} & \texttt{chow-lin-maxlog} & -- & sum & 0.0000 & 0.0000 & $2.6\times 10^{-14}$ & -- \\
\texttt{cl\_quilis} & \texttt{chow-lin-minrss-quilis} & truncated\_rho=-1 & sum & 0.9990 & 0.9990 & $1.1\times 10^{-5}$ & $3.4\times 10^{-13}$ \\
\texttt{dc\_add\_h0} & \texttt{denton-cholette} & criterion=additive, h=0 & sum & -- & -- & $2.7\times 10^{-10}$ & -- \\
\texttt{dc\_add\_h2} & \texttt{denton-cholette} & criterion=additive, h=2 & sum & -- & -- & $1.0\times 10^{-11}$ & -- \\
\texttt{dc\_noind} & \texttt{denton-cholette} & -- & sum & -- & -- & $6.2\times 10^{-15}$ & -- \\
\texttt{dc\_prop} & \texttt{denton-cholette} & criterion=proportional, h=1 & sum & -- & -- & $6.9\times 10^{-13}$ & -- \\
\texttt{denton\_add} & \texttt{denton} & criterion=additive, h=1 & sum & -- & -- & $4.4\times 10^{-11}$ & -- \\
\texttt{denton\_prop\_h2} & \texttt{denton} & criterion=proportional, h=2 & sum & -- & -- & $1.4\times 10^{-7}$ & -- \\
\texttt{dyn\_fixed} & \texttt{dynamic-fixed} & fixed\_rho=0.5 & sum & 0.5000 & 0.5000 & $3.0\times 10^{-9}$ & -- \\
\texttt{dyn\_maxlog} & \texttt{dynamic-maxlog} & -- & sum & 0.0000 & 0.0000 & $1.5\times 10^{-14}$ & -- \\
\texttt{fast} & \texttt{fast} & -- & sum & 1.0000 & 1.0000 & $4.9\times 10^{-11}$ & -- \\
\texttt{fernandez} & \texttt{fernandez} & -- & sum & 0.0000 & 0.0000 & $7.6\times 10^{-14}$ & -- \\
\texttt{fernandez\_noint} & \texttt{fernandez} & include\_intercept=False & sum & 0.0000 & 0.0000 & $8.8\times 10^{-14}$ & -- \\
\texttt{lit\_fixed} & \texttt{litterman-fixed} & fixed\_rho=0.7 & sum & 0.7000 & 0.7000 & $7.5\times 10^{-11}$ & -- \\
\texttt{lit\_maxlog} & \texttt{litterman-maxlog} & truncated\_rho=-1 & sum & -0.9990 & -0.9990 & $8.7\times 10^{-7}$ & $8.3\times 10^{-14}$ \\
\texttt{lit\_minrss} & \texttt{litterman-minrss} & truncated\_rho=-1 & sum & 0.9358 & 0.9358 & $8.7\times 10^{-5}$ & $1.1\times 10^{-8}$ \\
\texttt{ols} & \texttt{ols} & -- & sum & -- & 0.0000 & $1.5\times 10^{-14}$ & -- \\
\texttt{uniform} & \texttt{uniform} & -- & sum & -- & -- & $4.0\times 10^{-15}$ & -- \\
\texttt{denton1971\_add\_h1} & \texttt{denton} & variant=original, criterion=additive, h=1 & sum & -- & -- & $4.0\times 10^{-15}$ & -- \\
\texttt{denton1971\_dc\_add\_h1} & \texttt{denton} & variant=cholette, criterion=additive, h=1 & sum & -- & -- & $3.3\times 10^{-15}$ & -- \\
\texttt{denton1971\_prop\_h1} & \texttt{denton} & variant=original, criterion=proportional, h=1 & sum & -- & -- & $4.7\times 10^{-15}$ & -- \\
\bottomrule
\end{tabular}

%% file: tables/recovery_exports.tex
\begin{tabular}{lrrrr}
\toprule
Method & $\hat\rho$ & RMSE level & RMSE growth (pp) & corr.\ growth \\
\midrule
\texttt{chow-lin-maxlog} & 0.998 & 429.7 & 6.776 & 0.488 \\
\texttt{chow-lin-maxlog (no intercept)} & 0.999 & 429.7 & 6.776 & 0.488 \\
\texttt{fernandez} & 0.000 & 429.8 & 6.774 & 0.488 \\
\texttt{litterman-maxlog} & 0.504 & 432.8 & 6.799 & 0.485 \\
\texttt{dynamic-maxlog} & 0.914 & 467.8 & 7.491 & 0.287 \\
\texttt{proietti-maxlog} & 0.958 & 441.2 & 6.938 & 0.479 \\
\texttt{denton-cholette} & -- & 461.7 & 7.375 & 0.471 \\
\texttt{denton} & -- & 461.8 & 7.497 & 0.451 \\
\texttt{uniform} & -- & 361.9 & 5.828 & 0.696 \\
\bottomrule
\end{tabular}

%% file: tables/recovery_gdp.tex
\begin{tabular}{lrrrrr}
\toprule
Method & $\hat\rho$ & RMSE level & RMSE growth (pp) & corr.\ growth & RMSE 2009Q1--Q3 \\
\midrule
\texttt{chow-lin-maxlog} & 0.945 & 27.0 & 0.594 & 0.739 & 309.1 \\
\texttt{chow-lin-maxlog (no intercept)} & 0.992 & 27.0 & 0.596 & 0.739 & 293.9 \\
\texttt{fernandez} & 0.000 & 26.9 & 0.591 & 0.741 & 297.2 \\
\texttt{litterman-maxlog} & 0.000 & 26.9 & 0.591 & 0.741 & 297.2 \\
\texttt{dynamic-maxlog} & 0.000 & 31.4 & 0.653 & 0.701 & 410.3 \\
\texttt{proietti-maxlog} & 0.940 & 27.0 & 0.596 & 0.739 & 292.4 \\
\texttt{denton-cholette} & -- & 27.4 & 0.606 & 0.733 & 294.1 \\
\texttt{denton} & -- & 47.1 & 1.570 & 0.408 & 294.1 \\
\texttt{uniform} & -- & 79.6 & 1.680 & 0.306 & -- \\
\bottomrule
\end{tabular}

%% file: tables/mc_coverage.tex
\begin{tabular}{lrr}
\toprule
Setting & Prediction intervals (\%) & Slope intervals (\%) \\
\midrule
$\rho$ known (0.7) & 95.2 & 93.8 \\
$\rho$ estimated (maxlog) & 95.8 & 89.8 \\
\bottomrule
\end{tabular}

%% file: tables/mc_common_factor.tex
\begin{tabular}{rrr}
\toprule
$n_l$ & Size (\%) & Power (\%) \\
\midrule
20 & 7.0 & 100.0 \\
40 & 6.6 & 100.0 \\
80 & 5.6 & 100.0 \\
\bottomrule
\end{tabular}

%% file: tables/swiss_backtest.tex
\begin{tabular}{lrrrrr}
\toprule
Candidate & Origins & Bias & MAE & RMSE & MAPE (\%) \\
\midrule
Chow-Lin, exports & 5 & 11.21 & 57.75 & 63.66 & 5.89 \\
Chow-Lin, exports + imports & 5 & -5.06 & 62.15 & 70.79 & 6.31 \\
Fern\'andez, exports & 5 & 10.51 & 70.62 & 78.82 & 7.08 \\
Denton-Cholette, exports & 5 & 18.38 & 75.70 & 82.57 & 7.65 \\
\bottomrule
\end{tabular}

%% file: tables/swiss_conformity.tex
\begin{tabular}{lrrrrr}
\toprule
Method & corr.\ levels & corr.\ growth & sd growth (est.) & sd growth (ind.) & Volatility ratio \\
\midrule
\texttt{chow-lin-maxlog} & 0.998 & 0.956 & 6.85 & 7.87 & 0.871 \\
\texttt{fernandez} & 0.996 & 0.971 & 4.81 & 7.87 & 0.612 \\
\texttt{litterman-maxlog} & 0.996 & 0.971 & 4.81 & 7.87 & 0.612 \\
\texttt{denton-cholette} & 0.996 & 0.990 & 7.87 & 7.87 & 1.001 \\
\texttt{dynamic-maxlog} & 0.998 & 0.956 & 6.85 & 7.87 & 0.871 \\
\texttt{proietti-maxlog} & 0.997 & 0.940 & 10.10 & 7.87 & 1.284 \\
\bottomrule
\end{tabular}

%% file: tables/swiss_revisions.tex
\begin{tabular}{lrrrrr}
\toprule
Method & MAR level (\%) & MAR growth (pp) & MAR level, latest & MAR growth, latest & Max.\ level (\%) \\
\midrule
\texttt{denton-cholette} & 0.04 & 0.03 & 0.90 & 0.82 & 3.37 \\
\texttt{chow-lin-maxlog} & 0.25 & 0.34 & 0.50 & 0.60 & 2.42 \\
\texttt{fernandez} & 0.32 & 0.44 & 0.74 & 0.76 & 3.24 \\
\texttt{litterman-maxlog} & 0.33 & 0.44 & 0.75 & 0.77 & 3.24 \\
\texttt{dynamic-maxlog} & 0.80 & 1.12 & 0.92 & 1.38 & 6.68 \\
\bottomrule
\end{tabular}

%% file: tables/timing.tex
\begin{tabular}{lrrlr}
\toprule
Problem & $n_l$ & $n$ & Method & Median time (s) \\
\midrule
annual $\to$ quarterly & 40 & 160 & \texttt{chow-lin-maxlog} & 0.102 \\
annual $\to$ quarterly & 40 & 160 & \texttt{denton-cholette} & 0.003 \\
quarterly $\to$ monthly & 80 & 240 & \texttt{chow-lin-maxlog} & 0.235 \\
quarterly $\to$ monthly & 80 & 240 & \texttt{denton-cholette} & 0.003 \\
annual $\to$ monthly & 40 & 480 & \texttt{chow-lin-maxlog} & 0.143 \\
annual $\to$ monthly & 40 & 480 & \texttt{denton-cholette} & 0.004 \\
monthly $\to$ daily (30-day months) & 120 & 3600 & \texttt{chow-lin-maxlog} & 2.778 \\
monthly $\to$ daily (30-day months) & 120 & 3600 & \texttt{denton-cholette} & 0.010 \\
\bottomrule
\end{tabular}

%% file: sections/discussion.tex
\section{Discussion and conclusions}\label{sec:discussion}

\subsection{Limitations}

The prediction and coefficient intervals are conditional on $\rho$ and on
Gaussian residuals. They do not include the sampling variability of an
estimated $\rho$; in the simulation of Section~\ref{sec:montecarlo} this left
the coverage of the prediction intervals close to the nominal level but
reduced that of the slope intervals. Nor do they include the uncertainty about
the residual process or the choice of indicators, which in applications is
often the dominant source of error (Section~\ref{sec:recovery}).

All estimators are linear in the benchmarks and assume an additive model in
levels. Logarithmic and other non-linear specifications, state-space
formulations estimated with the Kalman filter
\citep{harvey1989forecasting,proietti2006temporal}, Bayesian and regularised
estimators for large sets of indicators are not provided. The dynamic Chow--Lin
and ADL models are experimental: in the experiments of
Section~\ref{sec:validation} the dynamic model was the least accurate
indicator-based method and revised most, the ADL model added volatility
relative to the indicator, and the ADL model has no reference implementation in
R against which it can be compared.

Systems of series are handled by the two-step approach: each series is
disaggregated and then reconciled with a contemporaneous total. Simultaneous
multivariate estimation, in which the series share information during
estimation, is not implemented. Padding of the sample with artificial
benchmarks is a heuristic whose effect depends on the data; the package records
the artificial values and counts them in the regression statistics, but cannot
assess whether the assumed continuation is correct.

Finally, every tool in Section~\ref{sec:assessment} is indirect. Backtests
measure the extrapolation of the aggregate, revision analysis measures
stability rather than accuracy, and conformity describes the relation with the
indicator without saying which relation is correct. Without a high-frequency
benchmark, the distribution within low-frequency periods cannot be validated
from the data, and the choice of indicator and residual model remains a
modelling decision.

\subsection{Conclusions}

\pkg{tempdisagg} provides the regression-based and benchmarking methods of
temporal disaggregation in Python with the formulas and defaults of the R
package \pkg{tempdisagg}, and verifies the equivalence automatically: for the
29 reference configurations, predictions and statistics agree to between
$10^{-15}$ and $10^{-7}$ in relative terms with $\rho$ fixed, and to about
$10^{-8}$ with $\rho$ estimated once the difference in optimiser tolerance is
removed. A single GLS engine based on the triangular factor of the residual
covariance makes the cost of a likelihood evaluation linear in the number of
high-frequency periods. Each fit returns complete inference and the
information needed to judge it: the profile of $\rho$ with identification
warnings, prediction variances, a record of every artificial value, and typed
warnings for statistical problems.

The experiments with known high-frequency series illustrate the main practical
point of the literature: the quality of a disaggregation depends more on the
indicator than on the estimator, and exact aggregation makes in-sample
comparisons uninformative. The package therefore treats assessment as part of
the workflow, with backtesting, revision analysis, conformity diagnostics and a
common-factor test. Possible extensions include intervals that account for the
estimation of $\rho$, simultaneous multivariate estimation, and state-space
formulations of the dynamic models.